\documentclass[11pt]{article}
\usepackage[margin=1in]{geometry}
\usepackage{graphicx}
\usepackage{subcaption}
\usepackage{parskip}
\usepackage{titlesec}
\usepackage{hyperref}
\usepackage{amsmath, amssymb}
\usepackage{float}
\usepackage{booktabs}
\usepackage{natbib}
\usepackage{xcolor}
\usepackage{authblk}

\title{How recombination rates affect escape from low-fitness states}
\author[1]{Elisa Heinrich-Mora}
\author[1]{Chase Van Amburg}
\author[1]{Marcus W. Feldman\thanks{Corresponding author: Marcus W. Feldman, mfeldman@stanford.edu}}

\affil[1]{Department of Biology, Stanford University, Stanford, CA 94305, USA}

\date{}

\begin{document}

\newcommand{\m}{\ensuremath{M_1}}
\newcommand{\M}{\ensuremath{M_2}}
\newcommand{\rmm}{\ensuremath{r_{11}}}
\newcommand{\rmM}{\ensuremath{r_{12}}}
\newcommand{\rMM}{\ensuremath{r_{22}}}

\maketitle

\begin{abstract}
 Adaptation often requires the assembly of favorable combinations of mutations that are individually deleterious. In such cases, populations may remain trapped at low-fitness genetic states even when a higher-fitness genotype exists. Recombination plays a dual role in this process because it can both generate and disrupt advantageous multilocus combinations. Previous work showed that the balance between selection and recombination determines whether populations cross fitness valleys or persist in low-fitness states associated with demographic decline. We study this problem in a three-locus model consisting of two selected loci and a recombination modifier locus. The modifier has no direct effect on fitness but alters the recombination rate between the selected loci, allowing recombination itself to evolve. We characterize the fixation states of the system and derive explicit conditions for the local stability of the low-fitness fixation set. Stability is determined by the interaction between selection strength, recombination among selected loci, recombination between the modifier and selected loci, and modifier composition. In the classical two-locus model, stability of the low-fitness state depends on a single recombination parameter. In contrast, the modifier model generates a continuum of fixation states whose stability varies with modifier frequency. We show that populations with identical selected-haplotype frequencies can differ in stability solely because they differ in modifier frequencies. Moreover, the relative magnitudes of modifier-dependent recombination rates determine whether modifier polymorphism stabilizes or destabilizes the low-fitness state. These results show that genetic variation affecting recombination changes evolutionary outcomes not only by altering the formation of favorable multilocus combinations, but also by changing the stability of alternative evolutionary states. Consequently, recombination modifiers can either reinforce persistence in low-fitness genetic configurations or facilitate escape toward high-fitness adaptive states.
\end{abstract}

\section*{Introduction}
When a population's environment changes rapidly, evolution of the genetic composition and demography of the population may occur on comparable timescales~\citep{yoshida2003rapid, hairston2005rapid, carroll2007evolution}. This occurs, for example, during abrupt environmental deterioration, antimicrobial treatment, or rapid climatic change~\citep{parmesan2006ecological, bell2009evolutionary, carlson2014evolutionary}. In such settings, population rescue---the recovery of a declining population through genetic or ecological intervention---depends not only on whether variants adapted to the new environment are available, but also on how rapidly they can increase before demographic decline becomes irreversible~\citep{gomulkiewicz1995evolution, orr2014population}. In such situations, transient evolutionary dynamics can be as important as properties of potential equilibria. If populations remain for extended periods in low-fitness states, interventions may not be able to prevent substantial declines in population size, or extinction, even when a higher-fitness equilibrium is theoretically possible~\citep{gomulkiewicz1995evolution, bovier2019crossing}.

For genetic intervention, an important case concerns adaptation that depends on interactions between mutations, which are deleterious separately, but advantageous~in combination \citep{crow1965evolution, eshel1970chance}. In this situation, no single mutational step increases fitness, but the multilocus combination does. This phenomenon, known as reciprocal sign epistasis, prevents the higher-fitness genotype from being reached through a sequence of fitness-increasing changes and instead requires passing through intermediate genotypes of lower fitness \citep{feldman1971equilibrium, weinreich2005sign, poelwijk2011reciprocal}. Selection initially acts against the intermediate states required to assemble the final advantageous combination. This problem is relevant in several evolutionary contexts, including the evolution of coadapted gene complexes and supergenes, the evolutionary consequences of recombination and sexual reproduction, and the dynamics of linkage disequilibrium ~\citep{lewontin1964interaction, bodmer1967linkage, feldman1972selection, barton1995general, burton1999genetic, devisser2014what}.

\cite{wirtz_impacts_2026} studied this problem in an eco-evolutionary framework in which multilocus genetic composition feeds back on population growth. They showed that recombination can either facilitate the assembly of a high-fitness genotype or break apart favorable multilocus combinations once they arise. This interaction determines whether a population's frequency configuration moves towards a high-fitness state or remain near lower-fitness states for extended periods. In particular, they identified parameter regimes in which the outcome depends on the initial genetic composition of the population. They further showed that chromosome fusions segregating at low initial frequency can accelerate recovery by reducing recombination between interacting loci.

Here we analyze the genetic dynamics underlying these outcomes by focusing on haplotype frequencies at two loci under selection and recombination. We consider a haploid population with two selected loci, $A$ and $B$, each with two alleles, yielding four haplotypes $A_1B_1$, $A_1B_2$, $A_2B_1$, and $A_2B_2$ whose frequencies  are $x_1,\ldots,x_4$, respectively, whose fitnesses are $w_1 = 1$, $w_2 = w_3 = 1-s$, and $w_4 = 1+s$, where $s>0$ measures selection against single-mutant haplotypes and in favor of the double mutant. This defines a fitness valley: in a population initially dominated by $A_1B_1$, reaching the higher-fitness haplotype $A_2B_2$ requires passing through the intermediate haplotypes $A_1B_2$ and $A_2B_1$, both of which have lower fitness than $A_1B_1$. Recombination between the loci is assumed to occur at rate $r_{11}$. The resulting selection--recombination recursions define a discrete-time dynamical system on the simplex of haplotype frequencies \cite{feldman1971equilibrium}.

The qualitative behavior of this system is determined by the relationship between $r_{11}$ and $s$. If \(r_{11} < \frac{s}{1+s},\) only fixation on the double-mutant haplotype $A_2B_2$ is locally stable and $A_2B_2$ from any positive initial frequency. By contrast, if \(r_{11} > \frac{s}{1+s},\) the system is bistable: both fixation of $A_1B_1$ and fixation of $A_2B_2$ are locally stable with their domains of attraction, separated by an unstable surface. Properties of this surface were studied by \cite{feldman1971equilibrium}. The evolutionary outcome depends on the initial composition of the population with respect to this surface, which determines whether the population remains near lower-fitness states that may increase the risk of demographic decline. This condition, originally identified by \cite{feldman1971equilibrium}, underlies the regimes described by \cite{wirtz_impacts_2026}.

We extend this framework by introducing a recombination modifier locus with alleles $M_1$ and $M_2$. Recombination between the selected loci depends on the modifier genotype, with rates $r_{11}$, $r_{12}$, and $r_{22}$ corresponding to genotypes $M_1M_1$, $M_1M_2$, and $M_2M_2$, respectively. Recombination between locus $B$ and the modifier locus occurs at rate $r$. The state of the system is therefore described by eight haplotype frequencies.

Let \[ \mathcal E_1 = \left\{ (\alpha,0,0,0,1-\alpha,0,0,0): 0\le\alpha\le1 \right\} \] and \[ \mathcal E_4 = \left\{ (0,0,0,\alpha,0,0,0,1-\alpha): 0\le\alpha\le1 \right\} \]

denote the fixation sets corresponding to fixation of $A_1B_1$ and $A_2B_2$, respectively. In these sets, the modifier is polymorphic. To determine local stability of these sets, we linearize the recursion equations near them. Linearization near $\mathcal E_1$ yields a matrix $\mathbf A_1(\alpha)$ describing the first-order growth of rare $A_2B_2$ haplotypes across modifier backgrounds. The fixation state $\mathcal E_1$ is locally stable if $(1+s)\rho\!\left(\mathbf A_1(\alpha)\right)<1,$ where $\rho(\mathbf A_1(\alpha))$ denotes the spectral radius of $\mathbf A_1(\alpha)$. Analogously, linearization near $\mathcal E_4$ yields a matrix $\mathbf A_4(\alpha)$ describing the first-order growth of rare $A_1B_1$ haplotypes. In the present model, $\mathcal E_4$ is locally stable for all admissible parameter values, i.e., $(1+s)\rho\!\left(\mathbf A_4(\alpha)\right)<1.$ Whenever $(1+s)\rho\!\left(\mathbf A_1(\alpha)\right)<1,$ both fixation sets are locally stable and the system is bistable. Trajectories may then converge either to $\mathcal E_1$ or to $\mathcal E_4$, depending on the initial state.

In the two-locus system, domains of attraction to the fixation states are determined by the balance between selection strength $s$ and $r_{11}$, the rate of recombination between selected loci. When both fixation states are stable, the long-term fixation state is determined by the initial haplotype frequencies. In the three-locus system, however, the dynamics are not fully specified by the selected-haplotype frequencies alone; they also depend on the recombination structure $(r, r_{11}, r_{12}, r_{22})$ and on the distribution of selected haplotypes across modifier backgrounds.

 % % % % % % %
 %           %
 % M O D E L %
 %           %
 % % % % % % %
\section{Model} \label{sec:modelsetup}
Consider a large haploid population with two diallelic loci, \(A\in\{A_1,A_2\}\) and \(B\in\{B_1,B_2\}\), generating four haplotypes that are under selection
\begin{equation}
1:(A_1B_1),\qquad
2:(A_1B_2),\qquad
3:(A_2B_1),\qquad
4:(A_2B_2).
\label{eq:haplotype_definitions}
\end{equation}
A third diallelic locus \(M\in\{M_1,M_2\}\) controls the recombination rate but has no direct effect on fitness. Let \(x_i\) and \(y_i\) denote the frequencies of haplotype \(i\) associated with modifier alleles \(M_1\) and \(M_2\), respectively, with $x=(x_1, x_2, x_3, x_4)$, $y=(y_1, y_2, y_3, y_4)$. The total frequency of haplotype \(i\) is
\begin{equation}
u_i=x_i+y_i,
\qquad
\sum_{i=1}^4 u_i=1.
\label{eq:haplotype_frequencies}
\end{equation}
Fitness depends only on the \(A/B\) haplotype and is parameterized by \(s\in(0,1)\):
\begin{equation}
w_1=1,\qquad
w_2=w_3=1-s,\qquad
w_4=1+s.
\label{eq:fitness_scheme}
\end{equation}
Thus \(A_2B_2\) is the fittest haplotype, while \(A_1B_2\) and \(A_2B_1\) have reduced fitness. The mean fitness is
\begin{equation}
\bar w
=
\sum_{i=1}^4 w_i u_i,
\label{eq:mean_fitness}
\end{equation}
 (see \cite{feldman1971equilibrium} and \cite{wirtz_impacts_2026}). The state variables and fitnesses are summarized in Table~\ref{tab:model_summary}.

\begin{table}[ht]
\centering
\caption{Selected haplotypes, modifier-associated frequencies, and viabilities.}
\label{tab:model_summary}
\begin{tabular}{ccccc}
\hline
Index \(i\) & 1 & 2 & 3 & 4 \\
\hline
Selected haplotype
& \((A_1B_1)\)
& \((A_1B_2)\)
& \((A_2B_1)\)
& \((A_2B_2)\) \\

Total frequency
& \(u_1\)
& \(u_2\)
& \(u_3\)
& \(u_4\) \\

Frequency on \(M_1\)
& \(x_1\)
& \(x_2\)
& \(x_3\)
& \(x_4\) \\

Frequency on \(M_2\)
& \(y_1\)
& \(y_2\)
& \(y_3\)
& \(y_4\) \\

Fitness
& \(1\)
& \(1-s\)
& \(1-s\)
& \(1+s\) \\
\hline
\end{tabular}
\end{table}
At each generation, viability selection is followed by random pairing, recombination, and haploid reproduction. After viability selection,
\begin{equation}
x_i^s=x_i\frac{w_i}{\bar w},
\qquad
y_i^s=y_i\frac{w_i}{\bar w}.
\label{eq:selection_step}
\end{equation}
The rate of recombination between loci \(A\) and \(B\) depends on modifier genotype:
\begin{equation}
(M_1M_1): r_{11},
\qquad
(M_1M_2): r_{12},
\qquad
(M_2M_2): r_{22}.
\label{eq:modifier_recombination}
\end{equation}
Recombination between loci \(B\) and \(M\) occurs at rate $r\in(0,\tfrac12)$, and $0<r; r_{11}, r_{12}, r_{22} < 1/2$. Then in the next generation, the haplotype frequencies are:
\begin{align}
x_1' &= x_1\frac{w_1}{\bar w}
- \frac{\rmm}{\bar w^2}(w_1w_4x_1x_4-w_2w_3x_2x_3) \notag\\
&\quad
- \frac{\rmM}{\bar w^2}
\left(
w_1x_1[w_3y_3+w_4y_4]
-
w_3x_3[w_1y_1+w_2y_2]
\right) \notag\\
&\quad
- \frac{r}{\bar w^2}
\left(
w_1x_1[w_2y_2+w_3y_3+w_4y_4]
-
w_1y_1[w_2x_2+w_3x_3+w_4x_4]
\right) \notag\\
&\quad
+ \frac{r\rmM}{\bar w^2}
\left(
w_1x_1[w_3y_3+w_4y_4]
+w_3y_3[w_1x_1+w_2x_2]
-w_1y_1[w_3x_3+w_4x_4]
-w_3x_3[w_1y_1+w_2y_2]
\right), \notag
\end{align}
\begin{align}
x_2' &= x_2\frac{w_2}{\bar w}
+ \frac{\rmm}{\bar w^2}(w_1w_4x_1x_4-w_2w_3x_2x_3) \notag\\
&\quad
- \frac{\rmM}{\bar w^2}
\left(
w_2x_2[w_3y_3+w_4y_4]
-
w_4x_4[w_1y_1+w_2y_2]
\right) \notag\\
&\quad
- \frac{r}{\bar w^2}
\left(
w_2x_2[w_1y_1+w_3y_3+w_4y_4]
-
w_2y_2[w_1x_1+w_3x_3+w_4x_4]
\right) \notag\\
&\quad
+ \frac{r\rmM}{\bar w^2}
\left(
w_2x_2[w_3y_3+w_4y_4]
+w_4y_4[w_1x_1+w_2x_2]
-w_2y_2[w_3x_3+w_4x_4]
-w_4x_4[w_1y_1+w_2y_2]
\right), \notag
\end{align}
\begin{align} 
x_3' &= x_3\frac{w_3}{\bar w}
+ \frac{\rmm}{\bar w^2}(w_1w_4x_1x_4-w_2w_3x_2x_3) \notag\\
&\quad
+ \frac{\rmM}{\bar w^2}
\left(
w_1x_1[w_3y_3+w_4y_4]
-
w_3x_3[w_1y_1+w_2y_2]
\right) \notag\\
&\quad
- \frac{r}{\bar w^2}
\left(
w_3x_3[w_1y_1+w_2y_2+w_4y_4]
-
w_3y_3[w_1x_1+w_2x_2+w_4x_4]
\right) \notag\\
&\quad
+ \frac{r\rmM}{\bar w^2}
\left(
w_3x_3[w_1y_1+w_2y_2]
+w_1y_1[w_3x_3+w_4x_4]
-w_3y_3[w_1x_1+w_2x_2]
-w_1x_1[w_3y_3+w_4y_4]
\right), \notag
\end{align}
\begin{align}
x_4' &= x_4\frac{w_4}{\bar w}
- \frac{\rmm}{\bar w^2}(w_1w_4x_1x_4-w_2w_3x_2x_3) \notag\\
&\quad
+ \frac{\rmM}{\bar w^2}
\left(
w_2x_2[w_3y_3+w_4y_4]
-
w_4x_4[w_1y_1+w_2y_2]
\right) \notag\\
&\quad
- \frac{r}{\bar w^2}
\left(
w_4x_4[w_1y_1+w_2y_2+w_3y_3]
-
w_4y_4[w_1x_1+w_2x_2+w_3x_3]
\right) \notag\\
&\quad
+ \frac{r\rmM}{\bar w^2}
\left(
w_4x_4[w_1y_1+w_2y_2]
+w_2y_2[w_3x_3+w_4x_4]
-w_4y_4[w_1x_1+w_2x_2]
-w_2x_2[w_3y_3+w_4y_4]
\right).
\label{eq:x_recursions}
\end{align}

The recursions for \(y_i'\) are:
\begin{align}
y_1' &= y_1\frac{w_1}{\bar w}
- \frac{\rMM}{\bar w^2}(w_1w_4y_1y_4-w_2w_3y_2y_3) \notag\\
&\quad
- \frac{\rmM}{\bar w^2}
\left(
w_1y_1[w_3x_3+w_4x_4]
-
w_3y_3[w_1x_1+w_2x_2]
\right) \notag\\
&\quad
- \frac{r}{\bar w^2}
\left(
w_1y_1[w_2x_2+w_3x_3+w_4x_4]
-
w_1x_1[w_2y_2+w_3y_3+w_4y_4]
\right) \notag\\
&\quad
+ \frac{r\rmM}{\bar w^2}
\left(
w_1y_1[w_3x_3+w_4x_4]
+w_3x_3[w_1y_1+w_2y_2]
-w_1x_1[w_3y_3+w_4y_4]
-w_3y_3[w_1x_1+w_2x_2]
\right), \notag
\end{align}
\begin{align}
y_2' &= y_2\frac{w_2}{\bar w}
+ \frac{\rMM}{\bar w^2}(w_1w_4y_1y_4-w_2w_3y_2y_3) \notag\\
&\quad
- \frac{\rmM}{\bar w^2}
\left(
w_2y_2[w_3x_3+w_4x_4]
-
w_4y_4[w_1x_1+w_2x_2]
\right) \notag\\
&\quad
- \frac{r}{\bar w^2}
\left(
w_2y_2[w_1x_1+w_3x_3+w_4x_4]
-
w_2x_2[w_1y_1+w_3y_3+w_4y_4]
\right) \notag\\
&\quad
+ \frac{r\rmM}{\bar w^2}
\left(
w_2y_2[w_3x_3+w_4x_4]
+w_4x_4[w_1y_1+w_2y_2]
-w_2x_2[w_3y_3+w_4y_4]
-w_4y_4[w_1x_1+w_2x_2]
\right), \notag
\end{align}
\begin{align}
y_3' &= y_3\frac{w_3}{\bar w}
+ \frac{\rMM}{\bar w^2}(w_1w_4y_1y_4-w_2w_3y_2y_3) \notag\\
&\quad
+ \frac{\rmM}{\bar w^2}
\left(
w_1y_1[w_3x_3+w_4x_4]
-
w_3y_3[w_1x_1+w_2x_2]
\right) \notag\\
&\quad
- \frac{r}{\bar w^2}
\left(
w_3y_3[w_1x_1+w_2x_2+w_4x_4]
-
w_3x_3[w_1y_1+w_2y_2+w_4y_4]
\right) \notag\\
&\quad
+ \frac{r\rmM}{\bar w^2}
\left(
w_3y_3[w_1x_1+w_2x_2]
+w_1x_1[w_3y_3+w_4y_4]
-w_3x_3[w_1y_1+w_2y_2]
-w_1y_1[w_3x_3+w_4x_4]
\right), \notag
\end{align}
\begin{align}
y_4' &= y_4\frac{w_4}{\bar w}
- \frac{\rMM}{\bar w^2}(w_1w_4y_1y_4-w_2w_3y_2y_3) \notag\\
&\quad
+ \frac{\rmM}{\bar w^2}
\left(
w_2y_2[w_3x_3+w_4x_4]
-
w_4y_4[w_1x_1+w_2x_2]
\right) \notag\\
&\quad
- \frac{r}{\bar w^2}
\left(
w_4y_4[w_1x_1+w_2x_2+w_3x_3]
-
w_4x_4[w_1y_1+w_2y_2+w_3y_3]
\right) \notag\\
&\quad
+ \frac{r\rmM}{\bar w^2}
\left(
w_4y_4[w_1x_1+w_2x_2]
+w_2x_2[w_3y_3+w_4y_4]
-w_4x_4[w_1y_1+w_2y_2]
-w_2y_2[w_3x_3+w_4x_4]
\right),
\label{eq:y_recursions}
\end{align}

where \(\bar w=\sum_{i=1}^4 (x_i+y_i)w_i\).

 % % % % % % % % % % % % % % % % % % % %
 %                                     %
 % T W O - L O C U S   D Y N A M I C S %
 %                                     %
 % % % % % % % % % % % % % % % % % % % %
\section{Two-Locus Dynamics} \label{sec:2locus}
We first analyze the selected two-locus subsystem where $M_1$ is fixed and $y_i\equiv0,\,\,i=1,\dots,4.$

Recombination between loci \(A\) and \(B\) occurs at rate $r_{11}\in(0,\tfrac12)$. Under the fitness conditions \eqref{eq:fitness_scheme}, the double mutant \(A_2B_2\) has the highest fitness, whereas the single-mutant haplotypes \(A_1B_2\) and \(A_2B_1\) are deleterious. The haplotype frequencies satisfy $x_i\ge0$, $\sum_{i=1}^4x_i=\sum_{i=1}^4u_i=1$. After viability selection,
\begin{equation}
x_i^s
=
\frac{w_i x_i}{\bar w},
\qquad
\bar w
=
\sum_{i=1}^4w_i x_i,
\label{eq:two_locus_selection}
\end{equation}
where \(\bar w\) denotes the mean fitness. Linkage disequilibrium after selection is
\begin{equation}
D^s
=
x_1^s x_4^s-x_2^s x_3^s,
\label{eq:two_locus_Ds}
\end{equation}
the recombination step gives
\begin{align}
x_1' &= x_1^s-r_{11}D^s, \notag\\
x_2' &= x_2^s+r_{11}D^s, \notag\\
x_3' &= x_3^s+r_{11}D^s, \notag\\
x_4' &= x_4^s-r_{11}D^s.
\label{eq:two_locus_recursion}
\end{align}

\subsection{Equilibrium Structure}
An equilibrium of the system \eqref{eq:two_locus_recursion} is a frequency vector $(x_1^*,x_2^*,x_3^*,x_4^*)$ satisfying
\[
x_i'=x_i,
\qquad
i=1,\dots,4.
\]
Substituting \eqref{eq:two_locus_selection} into \eqref{eq:two_locus_recursion} yields
\begin{align}
x_1(\bar w-w_1) &= -r_{11}\bar w D^s, \notag\\
x_2(\bar w-w_2) &= r_{11}\bar w D^s, \notag\\
x_3(\bar w-w_3) &= r_{11}\bar w D^s, \notag\\
x_4(\bar w-w_4) &= -r_{11}\bar w D^s.
\label{eq:two_locus_equilibrium_balance}
\end{align}
These equations express the balance between selection and recombination at equilibrium. Assume \(s\in(0,1)\), \(r_{11}>0\), and \(x_i\ge 0\), \(\sum_i x_i=1\). At equilibrium, write
\[
C=r_{11}\bar w D^s .
\]
Then at equilibrium 
\[
x_1(\bar w-w_1)=-C,\qquad
x_2(\bar w-w_2)=C,\qquad
x_3(\bar w-w_3)=C,\qquad
x_4(\bar w-w_4)=-C.
\]
First consider boundary equilibria. If some \(x_i=0\), the corresponding equation forces \(C=0\). Since \(r_{11}>0\) and \(\bar w>0\), this implies \(D^s=0\). Hence the equilibrium equations reduce to
\[
x_i(\bar w-w_i)=0,
\qquad i=1,\ldots,4.
\]
Thus every haplotype present at a boundary equilibrium must have fitness equal to \(\bar w\). Since
\[
w_1=1,\qquad w_2=w_3=1-s,\qquad w_4=1+s,
\]
the only possible polymorphic boundary support is the edge with haplotypes \(2\) and \(3\). But on this edge, if \(x_2>0\) and \(x_3>0\), then
\[
D^s=x_1^sx_4^s-x_2^sx_3^s=-x_2^sx_3^s<0,
\]
contradicting \(D^s=0\). The only boundary equilibria are
\[
E_1=(1,0,0,0),\quad
E_2=(0,1,0,0),\quad
E_3=(0,0,1,0),\quad
E_4=(0,0,0,1).
\]
Now consider an internal equilibrium, so \(x_i^*>0\) for all \(i\). Then \(C\neq 0\), because if \(C=0\), all present haplotypes would need equal fitness, which is impossible when all four haplotypes are present. From
\[
x_4^*(\bar w^*-(1+s))=-C
\]
and \(x_4^*>0\), we obtain \(C>0\). Hence
\[
x_1^*=\frac{C}{1-\bar w^*},\qquad
x_2^*=\frac{C}{\bar w^*-(1-s)},\qquad
x_3^*=\frac{C}{\bar w^*-(1-s)},\qquad
x_4^*=\frac{C}{1+s-\bar w^*}.
\]
Positivity therefore requires $1-s<\bar w^*<1.$ In particular, $x_2^*=x_3^*$, and $x_1^*+x_2^*+x_3^*+x_4^*=1$. \(C\) is determined once \(\bar w^*\) is known. Hence any internal equilibrium is uniquely determined by \(\bar w^*\). To determine \(\bar w^*\), we substitute the equilibrium frequencies into the definition of the linkage disequilibrium after selection, \(D^{s,*}
=
x_1^{s,*}x_4^{s,*}
-
x_2^{s,*}x_3^{s,*}.\) Using the fitness conditions we obtain
\[
D^{s,*}
=
\frac{
(1+s)x_1^*x_4^*
-
(1-s)^2x_2^*x_3^*
}{(\bar w^*)^2}.
\]
Substituting the expressions for \(x_1^*,x_2^*,x_3^*,x_4^*\) and simplifying yields a single equation for \(\bar w^*\),
\begin{equation}\label{eq:quadratic}
    q(\bar w^*)= 3(\bar w^*)^2 + \left[-r_{11}s+3r_{11}+2s-6\right]\bar w^* + r_{11}s^2+2r_{11}s-3r_{11}-s^2-2s+3=0.
\end{equation}

Thus, the problem of finding an internal equilibrium reduces to finding roots of the quadratic equation \(q(\bar w^*)=0\) in the admissible interval \(1-s<\bar w^*<1\). We now show that there is at most one such root. First,
\[
q(1-s)=-2r_{11}s(1-s)<0,
\]
and
\[
q(1)=s\left[r_{11}(1+s)-s\right]>0 \qquad \text{if } r_{11} > \frac{s}{1+s}.
\]

Therefore, $q(1-s)<0$,  $q(1)>0$ if $r_{11}> \frac{s}{1+s}$. Hence there is one root of \eqref{eq:quadratic} in $[1-s, w^*]$ if $r_{11}>\frac{s}{1+s}$. If $r<\frac{s}{1+s}$, there is no root of \eqref{eq:quadratic} (see \cite{feldman1971equilibrium}).

An internal equilibrium exists if and only if $r_{11}>\frac{s}{1+s}.$ At the threshold \(r_{11}=s/(1+s)\), the root occurs at \(\bar w^*=1\), which is the boundary point \(E_1\), not an internal equilibrium. Thus, for \(s\in(0,1)\) and \(r_{11}>0\), the complete equilibrium set is $\{E_1,E_2,E_3,E_4\}$ when $r_{11} \le s / (1+s)$ and $\{E_1,E_2,E_3,E_4,E_{\mathrm{int}}\}$ when $r_{11} > s/(1+s)$, where $E_{\text{int}}$ is the root of \eqref{eq:quadratic}. There are no additional boundary or internal equilibria. If \(r_{11}=0\), the whole edge
\[
\{(0,x_2,x_3,0):x_2+x_3=1\}
\]
is also an equilibrium, because haplotypes \(2\) and \(3\) have equal fitness and recombination is absent. We denote the recombination cutoff value $s/(1+s)$ by $r_c$.

\subsection{Stability Analysis}
We now determine which equilibria are locally stable by linearizing \eqref{eq:two_locus_recursion} near each equilibrium. The single-mutant equilibria \(E_2\) and \(E_3\) are unstable if $s>0$.

At $E_1=(1,0,0,0)$,  \(\bar w=1\), and rare single-mutant haplotypes have growth rate \(1-s<1\), so they disappear. The only possible successful invader is \(A_2B_2\). Linearization gives
\begin{equation}
x_4'=(1+s)(1-r_{11})x_4.
\label{eq:E1_linearization}
\end{equation}
Therefore
\begin{equation}
E_1 \text{ is locally stable}
\text{ if }
(1+s)(1-r_{11})<1
\iff
r_{11}>r_c.
\label{eq:E1_stability}
\end{equation}

At $E_4=(0,0,0,1)$, the growth rate of \(A_1B_1\) is
\begin{equation}
x_1'=
\frac{1-r_{11}}{1+s}x_1.
\label{eq:E4_linearization}
\end{equation}
Since \(0\le r_{11}\le 1/2\) and \(s>0\),
\[
\frac{1-r_{11}}{1+s}<1.
\]
Thus \(E_4\) is locally stable for all admissible parameter values. It remains to classify the internal equilibrium \(E_{\mathrm{int}}\), which exists only when $r_{11} > r_c$.

Linearizing on the simplex, using coordinates \((x_1,x_2,x_4)\) with $x_3=1-x_1-x_2-x_4$, shows that perturbations breaking the symmetry \(x_2=x_3\) have eigenvalue $\lambda_1=(1-s)/\bar w^*$. Since \(1-s<\bar w^*<1\), this eigenvalue satisfies $0<\lambda_1<1$.

The remaining two eigenvalues describe perturbations within the symmetric subspace \(x_2=x_3\). A direct calculation of the constrained Jacobian at \(E_{\mathrm{int}}\) shows that one of these eigenvalues is larger than one whenever $r_{11}>r_c$. Hence \(E_{\mathrm{int}}\) is unstable. More precisely, it is a saddle: it is stable in the surface $x_2 = x_3$, which separates the domains of attraction to \(E_1\) and \(E_4\).

\subsection{Basins of Attraction}
The long-term behavior of the system depends on the relationship between recombination and selection. From the stability analysis of \(E_1=(1,0,0,0)\), \(E_1\) is unstable when $r_{11} < r_c$, and in this case, no internal equilibrium exists and every trajectory converges to $E_4=(0,0,0,1).$ When $r_{11}>r_c$ both boundary equilibria \(E_1\) and \(E_4\) are locally stable. In addition, an internal equilibrium
\[
E_{\mathrm{int}}
=
(x_1^*,x_2^*,x_3^*,x_4^*)
\]
exists and is a saddle point, as shown in Section~\ref{sec:2locus}. The separatrix divides the simplex into two basins of attraction: trajectories starting on one side converge to \(E_1\), and trajectories starting on the other side converge to \(E_4\). This bistable regime is of particular interest because populations can become trapped at the low-fitness equilibrium \(E_1\) despite the presence of the fitter genotype \(A_2B_2\).

To visualize the geometry of the basins, we first consider the invariant slice
\[
x_2=x_3=\frac{1-x_1-x_4}{2}.
\]
This restriction corresponds to imposing symmetry between the two single-mutant haplotypes. Since the recursions preserve the equality \(x_2=x_3\), trajectories initialized on this slice remain on it for all subsequent generations. The slice therefore provides a two-dimensional representation of the dynamics while preserving the essential structure of the full system.

\begin{figure}
    \centering
    \includegraphics[width=0.8\linewidth]{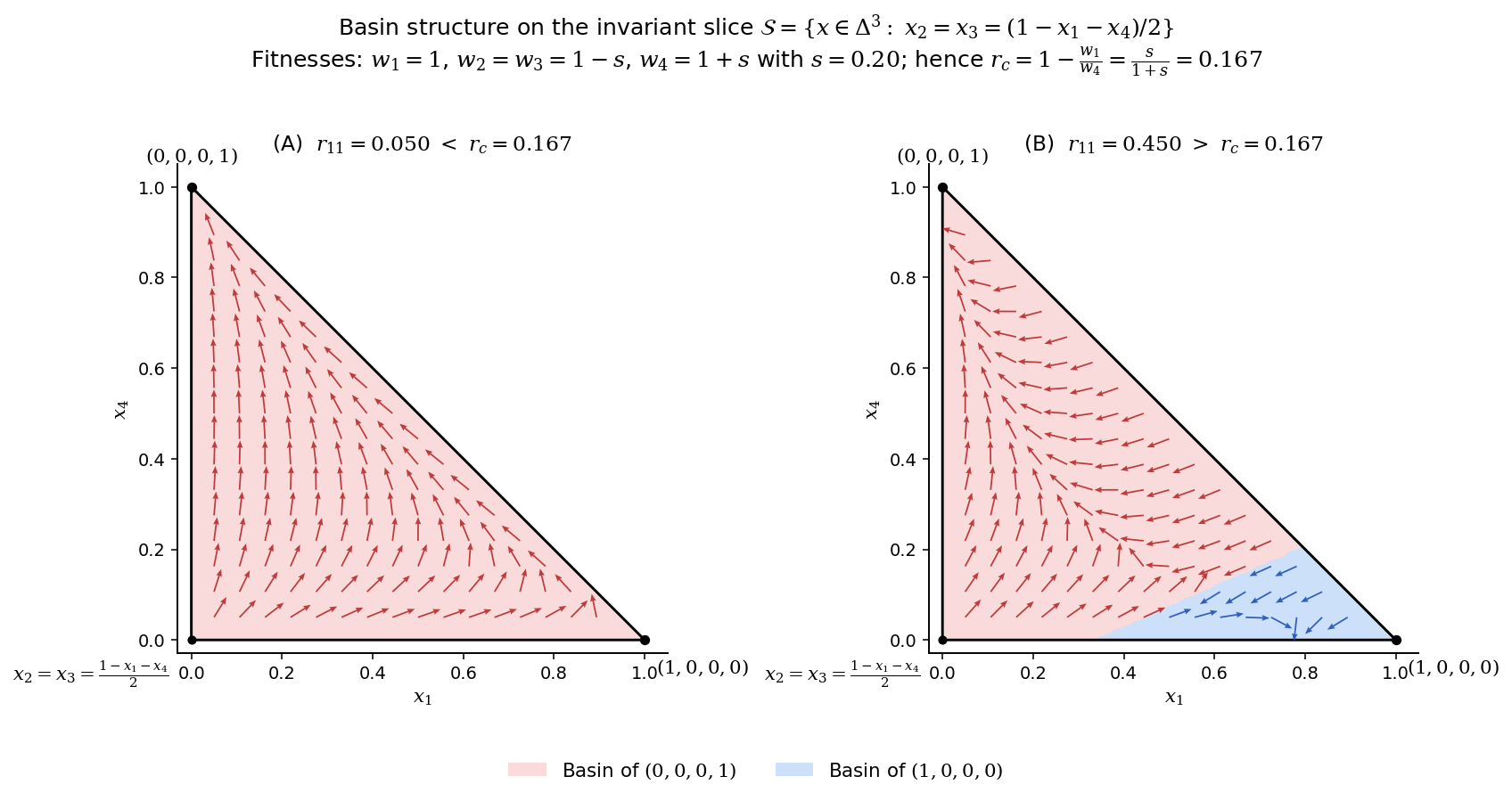}
\caption{\textbf{Dynamics with $x_2=x_3=\frac{1-x_1-x_4}{2}$} for \(w_1=1\), \(w_2=w_3=1-s\), and \(w_4=1+s\). Colors indicate numerically observed attraction regions and arrows indicate the direction of the dynamics. (A) \(r_{11}<r_c\): trajectories converge to \(E_4\). (B) \(r_{11}>r_c\): bistability, with trajectories converging either to \(E_1\) (blue) or to \(E_4\) (red).}
\label{fig:bassin_schematic}
\end{figure}

Figure~\ref{fig:bassin_schematic} illustrates the dynamics on this invariant slice. When \(r_{11}<r_c\), all trajectories converge to \(E_4\). When \(r_{11}>r_c\), the separatrix appears and partitions the state space into two domains of attraction. The location of this partition depends on the parameters \(r_{11}\) and \(s\), whereas the eventual outcome for a particular population depends on its initial haplotype frequencies.

The role of recombination can be understood directly from the stability condition. Since bistability requires $r_{11}>r_c,$ increasing \(r_{11}\) or decreasing \(s\) moves the system further into the bistable regime. As a consequence, the basin of attraction of \(E_1\) expands, increasing the set of initial conditions that converge to fixation of the lower-fitness haplotype \(A_1B_1\). The parameters determine the geometry of the basins, whereas the initial condition determines which basin contains the trajectory and therefore which equilibrium is reached.

\begin{figure}[H]
\centering
\includegraphics[width=\textwidth]{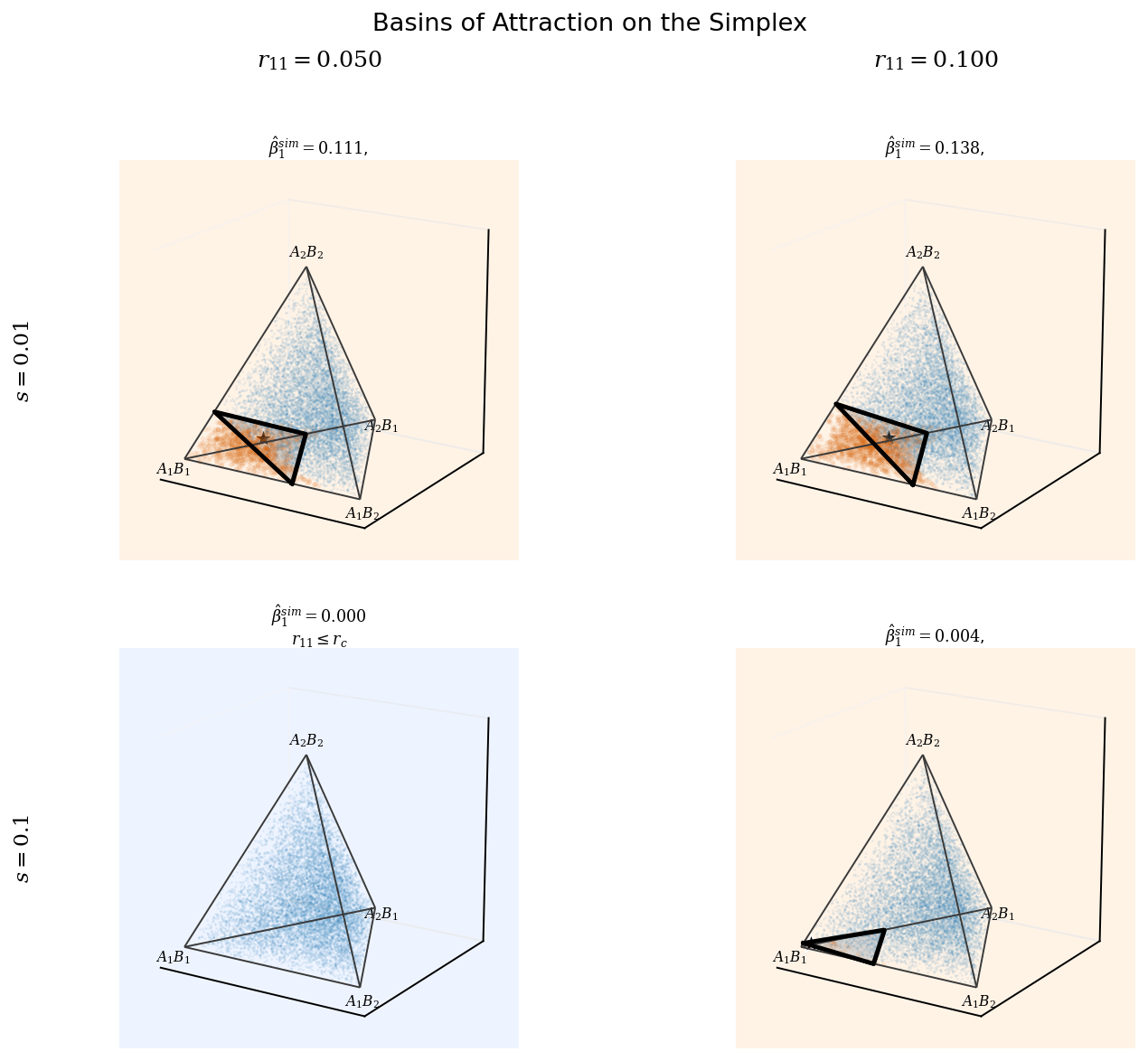}
\caption{\textbf{Basins of Attraction in the Simplex.} Each panel shows sampled initial conditions in the four-haplotype simplex for a fixed parameter pair \((s,r_{11})\). Orange points converge to \(E_1=(1,0,0,0)\), and blue points converge to \(E_4=(0,0,0,1)\) under iteration of the two-locus recursion. When
\(r_{11}>r_c\), the black polygon approximates the tangent-plane approximation obtained at the internal saddle equilibrium \(E_{\mathrm{int}}\), which is marked by a black star. Panel annotations report the simulated basin fraction \(\hat\beta_1^{\mathrm{sim}}\). When \(r_{11}\le r_c\), no internal equilibrium exists. [expain each block]}
\label{fig:simplex_outputs}
\end{figure}

To quantify basin size in the full simplex \(\Delta_3\), we sampled initial conditions
\[
\{x^{(k)}(0)\}_{k=1}^{N}
\subset \Delta_3
\]
and recorded the equilibrium reached by each trajectory. The basin fraction of \(E_1\) was estimated as
\begin{equation}
\hat{\beta}_1^{\mathrm{sim}}
=
\frac{
\#\left\{
k:\,
u^{(k)}(t)\rightarrow E_1
\right\}
}{N}.
\label{eq:beta_sim}
\end{equation}
Thus, \(\hat{\beta}_1^{\mathrm{sim}}\) estimates the proportion of the simplex occupied by the basin of attraction of \(E_1\). Figure~\ref{fig:simplex_outputs} shows representative simulations in the full simplex. Each point corresponds to a sampled initial condition and is colored according to its eventual equilibrium. Consistent with the invariant-slice analysis, the fraction of trajectories converging to \(E_1\) increases as recombination increases relative to selection. Initial conditions containing larger frequencies of the haplotype \(A_1B_1\) are also more likely to converge to \(E_1\). More generally, increasing \(r_{11}\) or decreasing \(s\) enlarges the volume of the simplex associated with the basin of attraction of the lower-fitness equilibrium.

Populations that remain for extended periods in low-fitness states may experience substantial demographic costs before adaptation occurs. To quantify this effect, for a tolerance \(\varepsilon>0\) define the convergence time
\begin{equation}
T_i^{\varepsilon}(x(0))
=
\inf
\left\{
t\ge 1:
\|x(t)-E_i\|_1
\le
\varepsilon
\right\},
\qquad
i\in\{1,4\}.
\label{eq:convergence_time}
\end{equation}
Thus, \(T_i^{\varepsilon}(x(0))\) is the number of generations required for a trajectory starting at \(x(0)\) to enter an \(\varepsilon\)-neighborhood of equilibrium \(E_i\).

\begin{figure}[t]
\centering
\includegraphics[width=\textwidth]{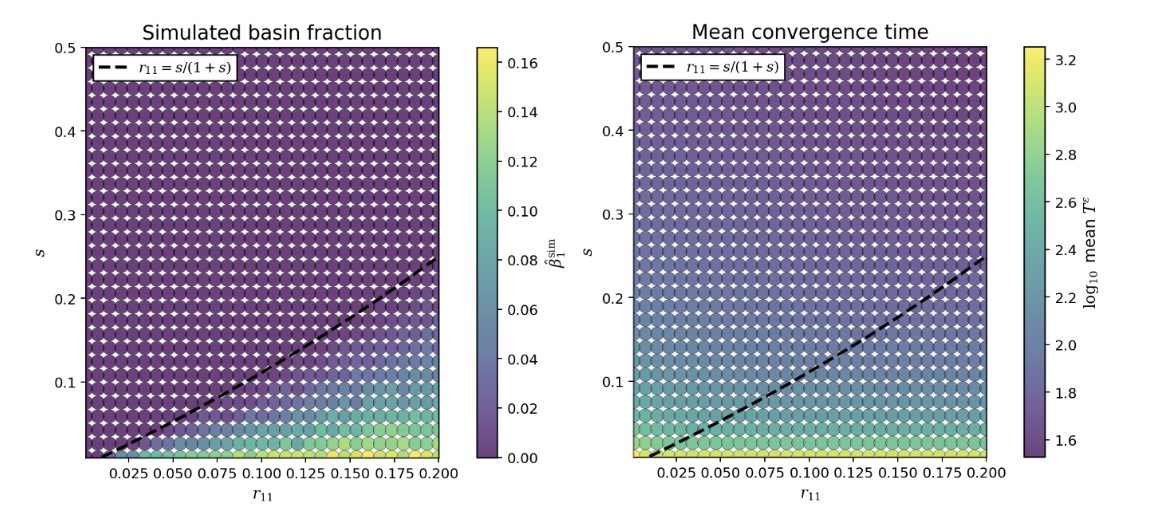}
\caption{
\textbf{Simulated basin fractions and time convergence in the \((r_{11},s)\)-parameter plane.} Left: simulated basin fraction \(\hat{\beta}_1^{\mathrm{sim}}\) obtained from sampled initial conditions in the simplex. Right: logarithm of the mean convergence time \(T^\varepsilon\). The dashed curve denotes the threshold \(r_{11}=s/(1+s)\).}
\label{fig:beta-plane-time}
\end{figure}

Figure~\ref{fig:beta-plane-time} summarizes how basin structure and convergence times vary across the \((r_{11},s)\) parameter space. The dashed curve $r_{11} = r_c$ separates the region, where all trajectories converge to \(E_4\), from the bistable region, where both \(E_1\) and \(E_4\) are locally stable. The left panel shows the simulated basin fraction \(\hat{\beta}_1^{\mathrm{sim}}\). Within the bistable region, the basin of attraction of \(E_1\) increases as recombination increases and selection decreases. Moving further above the threshold \(r_c\) shifts the stable manifold of the saddle equilibrium, increasing the volume of initial conditions that converge to \(E_1\). Stronger recombination more effectively breaks apart the favorable \(A_2B_2\) haplotype, while weaker selection reduces its fitness advantage. Both effects increase the probability that the population ultimately fixes the lower-fitness haplotype \(A_1B_1\).

The right panel shows the logarithm of the mean convergence time. In contrast to basin size, convergence times depend primarily on selection strength and only weakly on recombination. As \(s\) decreases, fitness differences among haplotypes become smaller, reducing the rate at which selection drives the population toward equilibrium. Consequently, trajectories can remain near lower-fitness states for many generations before approaching their final equilibrium. These results show that recombination primarily determines which equilibrium is reached, whereas selection primarily determines how quickly that equilibrium is approached.

\section{Three-Locus Dynamics} \label{sec:3locus}
We now return to the full three-locus system introduced in Section~\ref{sec:modelsetup}. The modifier locus \(M\) has no direct effect on viability, but alters the recombination rate between the selected loci \(A\) and \(B\). Selection on the modifier is therefore indirect: modifier alleles change in frequency only through their association with selected haplotypes.

The state of the population is \((x_1,x_2,x_3,x_4,y_1,y_2,y_3,y_4)\), where \(x_i\) and \(y_i\) denote the frequencies of selected haplotype \(i\) on modifier backgrounds \(M_1\) and \(M_2\), respectively. The state space is the simplex
\[
\Delta_7
=
\left\{
(x_1,\ldots,y_4)\in\mathbb R_{\ge0}^8:
\sum_{i=1}^4(x_i+y_i)=1
\right\}.
\]
The total frequency of selected haplotype \(i\in\{1,2,3,4\}\) is \(u_i=x_i+y_i\).
Fitness depends only on the selected haplotype, whereas recombination depends on modifier genotype. Recombination between \(A\) and \(B\) occurs at rates \(r_{11}\), \(r_{12}\), and \(r_{22}\) for modifier genotypes \(M_1M_1\), \(M_1M_2\), and \(M_2M_2\), respectively. Recombination between \(B\) and \(M\) occurs at rate \(r\).

\subsection{Equilibrium Structure} \label{equilibrium_three_locus}
The modifier locus changes the equilibrium structure even when the selected loci are fixed. In the two-locus model, fixation of a selected haplotype gives an isolated equilibrium. In the three-locus model, the same selected-locus fixation state does not determine the modifier frequency. Once a selected haplotype is fixed, the modifier is selectively neutral. Therefore, each two-locus fixation equilibrium becomes a one-dimensional equilibrium set. Let \(i\in\{1,2,3,4\}\), define
\begin{equation}
\mathcal E_i
=
\left\{
(x_1,\ldots,x_4,y_1,\ldots,y_4)\in\Delta_7:
x_i=\alpha,\;
y_i=1-\alpha,\;
x_j=y_j=0\;(j\neq i),\;
\alpha\in[0,1]
\right\}.
\label{eq:three_locus_fixation_sets}
\end{equation}
Here $\alpha$ is the frequency of modifier allele $M_1$ among individuals carrying selected haplotype $i$, while $(1-\alpha)$ is the corresponding frequency of the modifier allele $M_2$. Every point in \(\mathcal E_i\) is an equilibrium. To see this, suppose the population lies in  \(\mathcal E_i\). Then the marginal frequency of selected haplotype $i$ is $u_i = x_i +y_i$, $u_j=x_j+y_j=0$, where $j \neq i$, and the mean fitness is $\bar w=w_i$. Selection therefore leaves the state unchanged:
\[
x_i^s=x_i,
\qquad
y_i^s=y_i, 
\qquad 
x_j^s=y_j^s=0
\qquad (j \neq i).
\]
Moreover, all recombination terms vanish because every term involving another selected haplotype contains at least one absent frequency. The equilibrium sets  $\mathcal E_1, \mathcal E_2, \mathcal E_3, \mathcal E_4$, where $A_1B_1$, $A_1B_2$, $A_2B_1$ and $A_2B_2$, respectively, are fixed.  They are the direct analogues of the four fixation equilibria in the two-locus model, but now each contains a continuum of possible modifier frequencies. This distinction is important: the selected loci may be fixed, while the modifier locus remains polymorphic.

The full three-locus system may also admit additional equilibria, including equilibria with polymorphism at both the selected and modifier loci. We do not analyze such equilibria here. Their characterization requires solving the full nonlinear equilibrium system and is not necessary for our main objective. Instead, we focus on the fixation sets ($\mathcal E_i$), which extend the fixation equilibria of the two-locus model. These sets capture the adaptive endpoints of the selected loci and provide a natural framework for studying the effect of the modifier on evolutionary trajectories. Their stability governs the organization of the state space, the basins of attraction of alternative outcomes, and the long-term consequences of modifier evolution.

\subsection{Stability Analysis} \label{subsec:three_locus_stability}
The fixation sets $\mathcal E_i$  are not isolated equilibria: each point on $\mathcal E_i$ corresponds to a different modifier composition. 

First consider $\mathcal E_4 = (0, 0, 0, \alpha; 0, 0, 0, 1-\alpha)$, $\alpha \in [0,1]$, which corresponds to fixation of the highest-fitness selected haplotype $A_2B_2$. At any point of $\mathcal E_4$, $\bar w=w_4=1+s$, and small perturbations away from $\mathcal E_4$ are reduced by selection to first order. Recombination cannot increase their linear growth rate above that of the resident fittest haplotype. Thus $\mathcal E_4$ is locally stable. 

The fixation set $\mathcal E_1 = (\alpha,0,0,0;\,1-\alpha,0,0,0)$, requires a separate analysis. Here the resident selected haplotype is \(A_1B_1\), and since \(A_1B_2\) and \(A_2B_1\) have lower fitness than \(A_1B_1\), they do not produce instability. The relevant perturbation is the small frequency of the fitter haplotype \(A_2B_2\).

Fix a point \(\mathcal E_1(\alpha)\in\mathcal E_1\). Linearizing the recursion around this point and retaining the first-order terms in the small frequencies \(x_4\) and \(y_4\) gives \begin{equation} \begin{pmatrix} x_4'\\[0.3em] y_4' \end{pmatrix} = (1+s)\mathbf A(\alpha) \begin{pmatrix} x_4\\[0.3em] y_4 \end{pmatrix}, \label{eq:E1_linearization_three_locus} \end{equation} where \begin{equation} \mathbf A(\alpha) = \begin{pmatrix} 1-\alpha r_{11}-(1-\alpha)r_{12} -(1-\alpha)r(1-r_{12}) & \alpha r(1-r_{12}) \\[0.5em] (1-\alpha)r(1-r_{12}) & 1-(1-\alpha)r_{22}-\alpha r_{12} -\alpha r(1-r_{12}) \end{pmatrix}. \label{eq:invasion_matrix} \end{equation}
The entries of \(\mathbf A(\alpha)\) describe the first-order dynamics of rare \(A_2B_2\) haplotypes on the two modifier alleles. Because \(\mathbf A(\alpha)\) is nonnegative, the asymptotic growth rate of the rare \(A_2B_2\) perturbation is determined by the dominant eigenvalue of \((1+s)\mathbf A(\alpha)\). Therefore, 
\begin{equation} 
\mathcal E_1(\alpha) \text{ is locally stable} \quad \text{if}\quad \rho\!\left((1+s)\mathbf A(\alpha)\right)<1, \label{eq:E1_stability_condition} 
\end{equation} 
where \(\rho(\cdot)\) denotes the spectral radius. This condition shows that stability depends on the modifier composition \(\alpha\). To connect this result to the two-locus model, consider the endpoint \(\alpha=1\), where the population is fixed for modifier background \(M_1\). Then 
\[ 
\mathbf A(1) = \begin{pmatrix} 1-r_{11} & r(1-r_{12})\\ 0 & 1-r_{12}-r(1-r_{12}) \end{pmatrix}. 
\] 
The eigenvalues are the diagonal entries, and the dominant eigenvalue gives \[ \rho\!\left((1+s)\mathbf A(1)\right) = (1+s)(1-r_{11}). \] Thus \eqref{eq:E1_stability_condition} becomes \( r_{11}>r_c \), and the two-locus result is recovered when the modifier is fixed. In the three-locus model, however, the same selected-locus fixation state is a continuum of equilibria, and its stability can vary with \(\alpha\).

\subsection{Stability Conditions for \(\mathcal E_1\)}
We now determine when the \(A_1B_1\) fixation set \(\mathcal E_1\) is locally stable. In the three-locus model, rare \(A_2B_2\) haplotypes can occur on either modifier background. Their growth therefore depends on selection, recombination, and the modifier composition \(\alpha\) of the resident population. Fix a point \(\mathcal E_1(\alpha)\). From \eqref{eq:E1_stability_condition}, $\mathcal E_1(\alpha)$ is locally stable if $\rho\!\left((1+s)\mathbf A(\alpha)\right)<1$, namely \[\rho\!\left(\mathbf A(\alpha)\right) < \frac{1}{1+s}.\] Thus the dominant eigenvalue of \(\mathbf A(\alpha)\) must lie below the threshold \(1/(1+s)\). Write \[ \mathbf A(\alpha) = \begin{pmatrix} a & b\\ c & d \end{pmatrix}, \] where \[ a= 1-\alpha r_{11}-(1-\alpha)r_{12} -(1-\alpha)r(1-r_{12}), \] \[ b=\alpha r(1-r_{12}), \qquad c=(1-\alpha)r(1-r_{12}), \] and \[ d= 1-(1-\alpha)r_{22}-\alpha r_{12} -\alpha r(1-r_{12}). \]
Because \(\mathbf A(\alpha)\) is nonnegative, its dominant eigenvalue is its Perron root. Therefore, local stability is determined by the larger eigenvalue of this \(2\times2\) matrix. Let \[ p(\lambda) = \det(\lambda I-\mathbf A) = \lambda^2-\operatorname{tr}(\mathbf A)\lambda+\det(\mathbf A) \] be the characteristic polynomial. Its roots are the eigenvalues of \(\mathbf A\), and \(p\) is an upward-opening quadratic. Hence \[ \rho(\mathbf A)<\frac{1}{1+s} \] is equivalent to requiring that \(1/(1+s)\) lies to the right of both roots. For a \(2\times2\) nonnegative matrix this is equivalent to \[ \frac{1}{1+s}-a>0, \qquad \frac{1}{1+s}-d>0, \qquad \det\!\left(\frac{1}{1+s}I-\mathbf A\right)>0. \]
We now expand these three conditions. First, \[ \frac{1}{1+s}-a = \alpha r_{11} +(1-\alpha)r_{12} +(1-\alpha)r(1-r_{12}) -\frac{s}{1+s}. \] Second, \[ \frac{1}{1+s}-d = (1-\alpha)r_{22} +\alpha r_{12} +\alpha r(1-r_{12}) -\frac{s}{1+s}. \] Finally, \[ \det\!\left(\frac{1}{1+s}I-\mathbf A\right) = \left(\frac{1}{1+s}-a\right) \left(\frac{1}{1+s}-d\right)-bc, \] with \[ bc=\alpha(1-\alpha)r^2(1-r_{12})^2. \]

Therefore, \(\mathcal E_1(\alpha)\) is locally stable if and only if \[ \alpha r_{11} +(1-\alpha)r_{12} +(1-\alpha)r(1-r_{12}) > \frac{s}{1+s}, \] \[ (1-\alpha)r_{22} +\alpha r_{12} +\alpha r(1-r_{12}) > \frac{s}{1+s}, \] and \[ \left[ \alpha r_{11} +(1-\alpha)r_{12} +(1-\alpha)r(1-r_{12}) -\frac{s}{1+s} \right] \left[ (1-\alpha)r_{22} +\alpha r_{12} +\alpha r(1-r_{12}) -\frac{s}{1+s} \right] > \alpha(1-\alpha)r^2(1-r_{12})^2. \]

The first two inequalities require the effective recombination loss of rare \(A_2B_2\) haplotypes to exceed the selective threshold \(s/(1+s)\) on each modifier background. The third inequality is the coupling condition: the product of the two net losses must exceed the rate at which recombination moves rare \(A_2B_2\) haplotypes between modifier alleles.

Thus, stability depends on the resident modifier composition \(\alpha\), the recombination rates associated with each modifier background, and the coupling between backgrounds. As a check, if \(\alpha=1\), then \(c=0\) and the condition on the first modifier allele reduces to \( r_{11}>r_c, \) recovering the two-locus stability condition.

\paragraph{Parameter Effects. } The stability conditions can be written as \[ X(\alpha)>0,\qquad Y(\alpha)>0,\qquad X(\alpha)Y(\alpha)>Z(\alpha), \] where \[ X(\alpha) = \alpha r_{11} +(1-\alpha)r_{12} +(1-\alpha)r(1-r_{12}) -\frac{s}{1+s}, \] \[ Y(\alpha) = (1-\alpha)r_{22} +\alpha r_{12} +\alpha r(1-r_{12}) -\frac{s}{1+s}, \] and \[ Z(\alpha) = \alpha(1-\alpha)r^2(1-r_{12})^2. \]
The terms \(X\) and \(Y\) are the net first-order losses of rare \(A_2B_2\) haplotypes on the two modifier alleles, after subtracting the selective threshold \(s/(1+s)\). The term \(Z\) measures coupling between modifier alleles. Thus stability requires positive net loss on both alleles and requires the product of those losses to exceed the coupling term.

\begin{itemize}
    \item \textit{Selection} is destabilizing. Since \[ \frac{d}{ds}\left(\frac{s}{1+s}\right) = \frac{1}{(1+s)^2}>0, \] increasing \(s\) decreases both \(X\) and \(Y\), while leaving \(Z\) unchanged. Stronger selection for \(A_2B_2\) therefore makes \(\mathcal E_1(\alpha)\) harder to stabilize.
    \item \textit{Recombination rate $r_{11}$.} For \(r_{11}\), \[ \frac{\partial X}{\partial r_{11}}=\alpha, \qquad \frac{\partial Y}{\partial r_{11}}=0, \qquad \frac{\partial Z}{\partial r_{11}}=0. \] Thus \(r_{11}\) increases stability only through the \(M_1\) background, with an effect proportional to the resident frequency \(\alpha\). 
    \item \textit{Recombination rate $r_{22}$}. Similarly, \[ \frac{\partial Y}{\partial r_{22}}=1-\alpha, \qquad \frac{\partial X}{\partial r_{22}}=0, \qquad \frac{\partial Z}{\partial r_{22}}=0. \] Thus \(r_{22}\) increases stability only through the \(M_2\) background, with an effect proportional to \(1-\alpha\).
    \item \textit{Recombination rate $r_{12}$.} It affects both net losses and the coupling term: \[ \frac{\partial X}{\partial r_{12}}=(1-\alpha)(1-r), \qquad \frac{\partial Y}{\partial r_{12}}=\alpha(1-r), \] and \[ \frac{\partial Z}{\partial r_{12}} = -2\alpha(1-\alpha)r^2(1-r_{12}). \] Under the usual recombination range \(0\le r\le 1/2\), increasing \(r_{12}\) increases both \(X\) and \(Y\) and decreases \(Z\). Therefore \(r_{12}\) has a monotone stabilizing effect on \(\mathcal E_1(\alpha)\).
    \item \textit{Background recombination rate \(r\)}. It has two opposing effects: \[ \frac{\partial X}{\partial r}=(1-\alpha)(1-r_{12}), \qquad \frac{\partial Y}{\partial r}=\alpha(1-r_{12}), \] and \[ \frac{\partial Z}{\partial r} = 2\alpha(1-\alpha)r(1-r_{12})^2. \] Increasing \(r\) raises both net-loss terms, which favors stability. It also increases \(Z\), which strengthens coupling between modifier alleles and can oppose stability. Thus the net effect of \(r\) cannot be inferred from monotonicity alone; it must be evaluated from the full condition \[ X(\alpha)Y(\alpha)>Z(\alpha). \]
    \item \textit{Modifier allele frequency $\alpha$.} It changes both the allele-specific losses and the coupling term: \[ \frac{\partial X}{\partial \alpha} = r_{11}-r_{12}-r(1-r_{12}), \] \[ \frac{\partial Y}{\partial \alpha} = -r_{22}+r_{12}+r(1-r_{12}), \] and \[ \frac{\partial Z}{\partial \alpha} = (1-2\alpha)r^2(1-r_{12})^2. \]
    Increasing \(\alpha\) shifts weight from the \(M_2\) background to the \(M_1\) background. It increases \(X\) when \[ r_{11}>r_{12}+r(1-r_{12}), \] and decreases \(X\) otherwise. It increases \(Y\) when \[ r_{12}+r(1-r_{12})>r_{22}, \] and decreases \(Y\) otherwise. The coupling term \(Z\) is maximal at \(\alpha=1/2\) and vanishes at \(\alpha=0\) and \(\alpha=1\). Thus coupling is strongest when both modifier alleles are present and disappears when either modifier background is fixed.
\end{itemize}

\begin{figure}[htbp] 
\centering 
\includegraphics[width=0.75\textwidth]{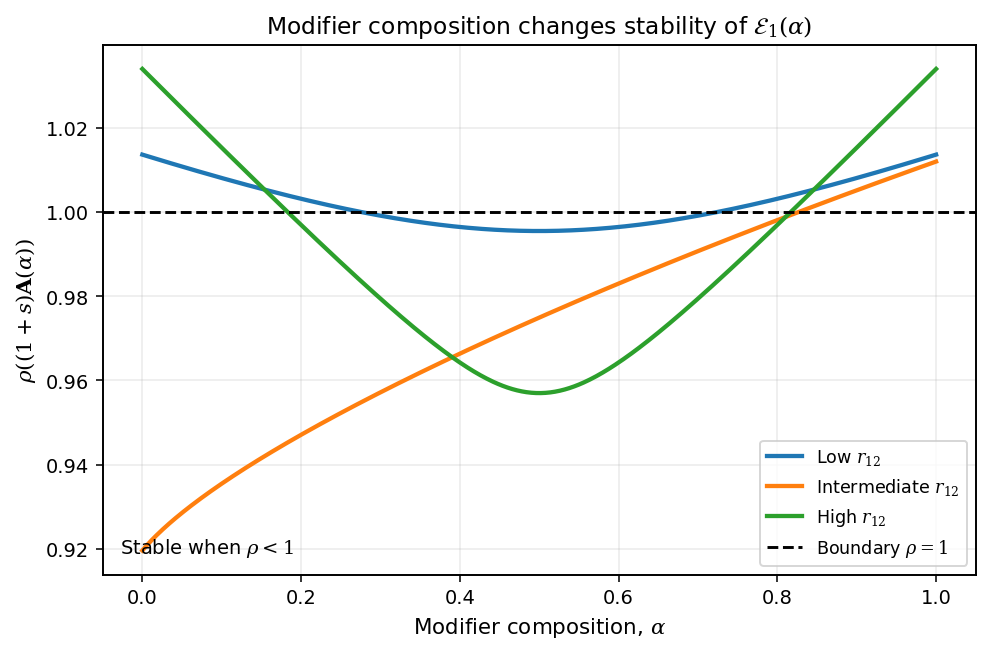} 
\caption{\textbf{Dependence of \(\mathcal E_1(\alpha)\) stability on modifier frequencies.}
The dominant eigenvalue \(\rho((1+s)\mathbf A_1(\alpha))\) is shown as a function of \(\alpha\), the frequency of $A_1B_1M_1$, where \(\mathbf A_1(\alpha)\) is the linearized invasion matrix for rare \(A_2B_2\) haplotypes near \(\mathcal E_1(\alpha)\). The fixation set \(\mathcal E_1(\alpha)\) is locally stable when \(\rho((1+s)\mathbf A_1(\alpha))<1\), corresponding to values below the dashed line. For all three curves, \(s=0.10\) and \(r=0.05\). The low-\(r_{12}\) case uses \((r_{11},r_{12},r_{22})=(0.16,0.03,0.16)\); the intermediate-\(r_{12}\) case uses \((0.08,0.12,0.18)\); and the high-\(r_{12}\) case uses \((0.06,0.20,0.06)\). These examples show that modifier allele frequencies can alter the local stability of \(\mathcal E_1(\alpha)\).}
\label{fig:E1_stability_alpha}
\end{figure}

These analytical results imply that local stability depends not only on the recombination rates themselves, but also on how modifier alleles are distributed in the population. Figure~\ref{fig:E1_stability_alpha} illustrates this dependence. For fixed values of \((r_{11},r_{12},r_{22},r,s)\), changing only the modifier composition can move the dominant eigenvalue across the stability threshold \[ \rho\!\left((1+s)\mathbf A(\alpha)\right)=1. \] Thus two populations fixed for the same selected haplotype \(A_1B_1\) can differ in stability solely because they differ in modifier composition.

\begin{figure}[htbp] 
\centering 
\includegraphics[width=\textwidth]{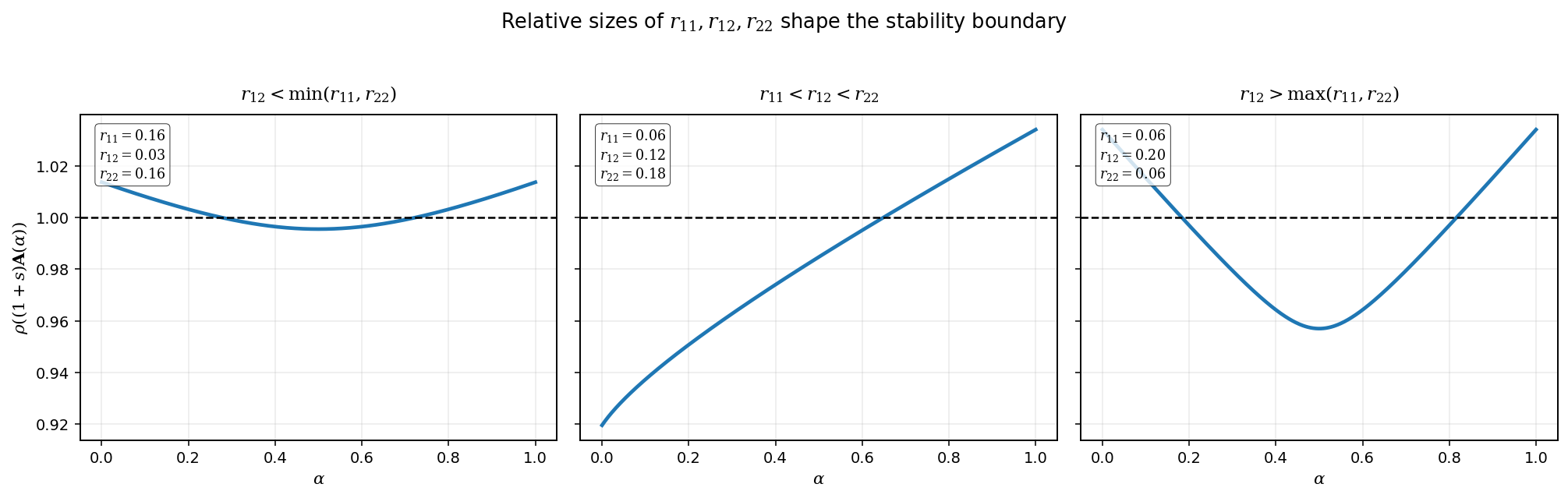} 
\caption{\textbf{Effect of relative recombination rates on stability.} Each panel shows \(\rho((1+s)\mathbf A_1(\alpha))\) as a frequency \(\alpha\) of modifier allele $M_1$, where stability requires values below the dashed line at \(\rho=1\). In all panels, \(s=0.10\) and \(r=0.05\). The left panel uses \((r_{11},r_{12},r_{22})=(0.16,0.03,0.16)\), so that \(r_{12}<\min(r_{11},r_{22})\). The middle panel uses \((r_{11},r_{12},r_{22})=(0.06,0.12,0.18)\), so that \(r_{11}<r_{12}<r_{22}\). The right panel uses \((r_{11},r_{12},r_{22})=(0.06,0.20,0.06)\), so that \(r_{12}>\max(r_{11},r_{22})\). These cases show that the ordering of modifiers' recombination rates can change how modifier frequency affects the local stability of \(\mathcal E_1(\alpha)\).}
\label{fig:E1_r_relationship}

\end{figure}

The shape of the stability profile depends on the relative magnitudes of \(r_{11}\), \(r_{12}\), and \(r_{22}\). Figure~\ref{fig:E1_r_relationship} shows three representative cases. If \[ r_{12}<\min\{r_{11},r_{22}\}, \] then mixed modifier backgrounds $(M_1M_2)$ have the lowest recombination rate, so intermediate modifier frequencies tend to reduce the disruption of rare \(A_2B_2\) haplotypes. If \[ r_{11}<r_{12}<r_{22}, \] the two backgrounds contribute asymmetrically, and stability changes as the population shifts from one modifier background to the other. If \[ r_{12}>\max\{r_{11},r_{22}\}, \] mixed modifier backgrounds have the highest recombination rate, so intermediate modifier frequencies tend to increase the disruption of rare \(A_2B_2\) haplotypes. Thus the effect of modifier polymorphism depends on the relative ordering of the recombination rates.

\begin{figure}[htbp] 
\centering 
\includegraphics[width=\textwidth]{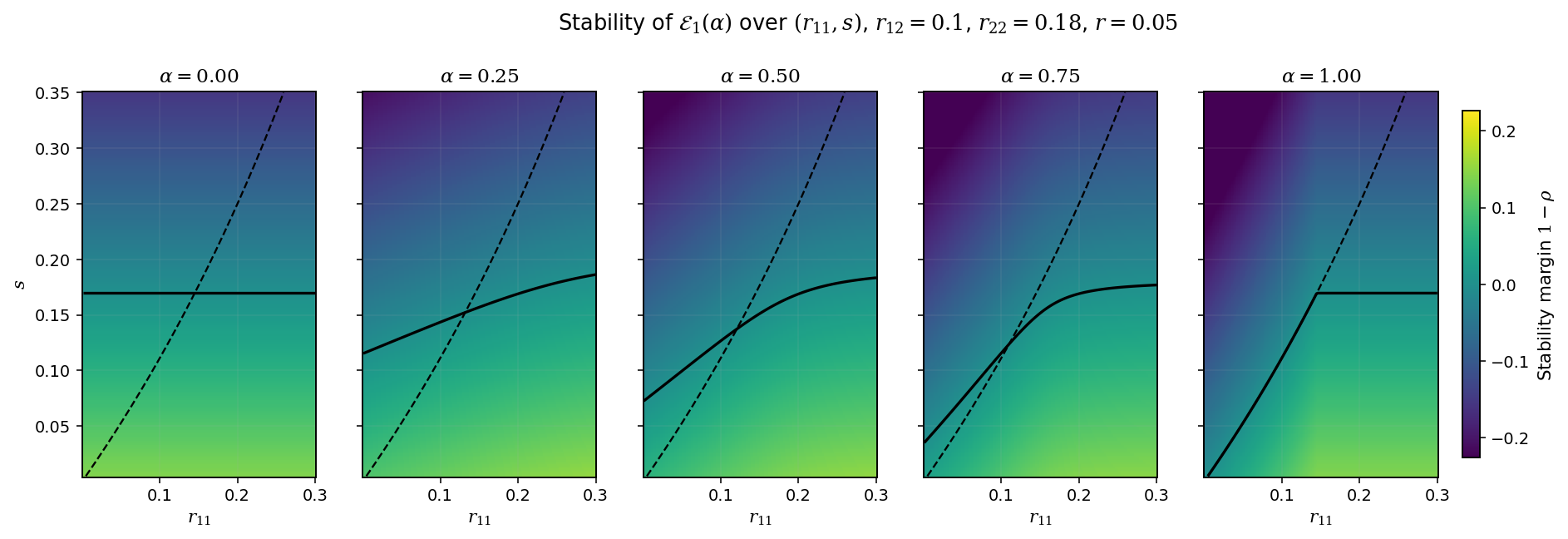} 
\caption{\textbf{Comparison of two-locus and three-locus stability boundaries.} Color indicates the stability margin \(1-\rho((1+s)\mathbf A(\alpha))\). Positive values correspond to local stability of \(\mathcal E_1(\alpha)\). The dashed curve is the two-locus threshold \(r_{11}=r_c\), while the solid contour is the exact three-locus stability boundary. Changing the modifier frequencies shifts both the location and shape of the stability region.} \label{fig:E1_boundary_comparison} 
\end{figure}

At the endpoints of the equilibrium set, the three-locus conditions reduce to simple thresholds. When \(\alpha=1\), 
\[ 
r_{11}>r_c, \qquad r+r_{12}-rr_{12}>r_c. 
\] 
When \(\alpha=0\), 
\[ 
r_{22}>r_c, \qquad r+r_{12}-rr_{12}>r_c. 
\] The first inequality in each case is the corresponding two-locus condition on the resident modifier background. The second condition reflects the possibility that rare \(A_2B_2\) haplotypes occur on the alternative modifier background. Figure~\ref{fig:E1_boundary_comparison} compares the three-locus stability boundary with the two-locus prediction in the \((r_{11},s)\) plane. The dashed curve is the two-locus threshold \( r_{11}=r_c \), whereas the solid curve is the exact three-locus boundary. As \(\alpha\) changes, the boundary shifts in both position and shape. Modifier composition therefore changes not only whether \(\mathcal E_1(\alpha)\) is stable, but also how stability depends on selection and recombination.

Overall, the modifier transforms the single two-locus threshold into a family of stability conditions indexed by \(\alpha\). There is still a requirement for a balance between selection and recombination: recombination must disrupt rare \(A_2B_2\) haplotypes faster than selection amplifies them. In the three-locus model, however, this balance depends on the modifier alleles of the rare haplotype and on recombination-mediated movement between backgrounds. Consequently, modifier polymorphism can either stabilize or destabilize the lower-fitness fixation set \(\mathcal E_1\), even when the selective regime is unchanged.

\subsection{Effect of recombination rates on evolution of modifier frequency}

\begin{figure}[H] 
\centering 
\includegraphics[width=0.8\textwidth]{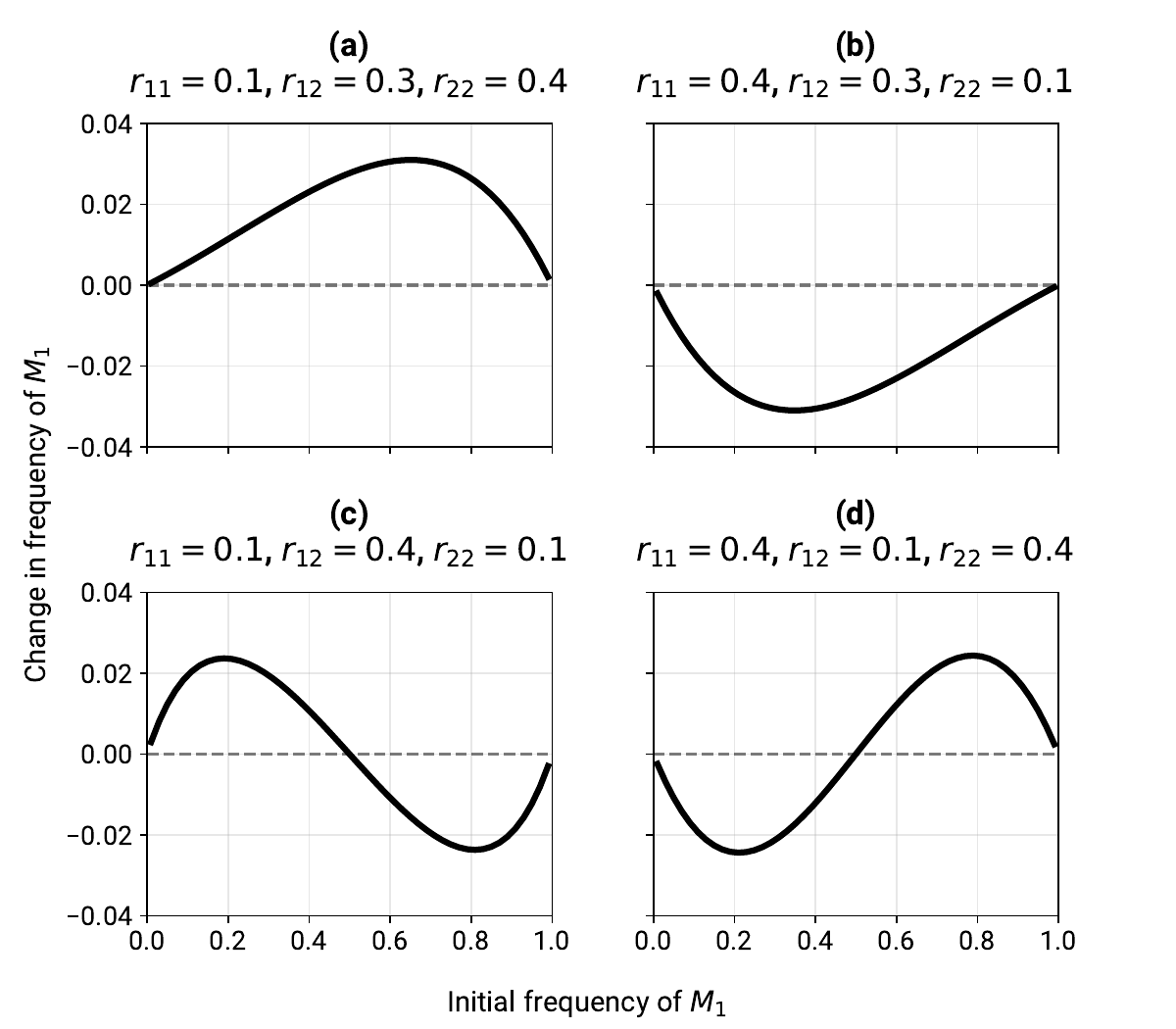} 
\caption{\textbf{\boldmath Change in modifier frequency from initial value to equilibrium for different recombination rates}. Trajectories run for 1000 generations with $r=0$ and $s=0.1$, and initialized at $u_0=(0.25, 0.25, 0.25, 0.25)$, where $x_0=\gamma u_0$ and $y_0 = (1-\gamma)u_0$, and $\gamma$ is the initial fraction of $M_1$. Other recombination parameters are (a) $r_{11}=0.1$, $r_{12} = 0.3$, and $r_{22}=0.4$, (b) $r_{11}=0.4$, $r_{12} = 0.3$, and $r_{22}=0.1$, (c) $r_{11}=0.4$, $r_{12} = 0.1$, and $r_{22}=0.4$, and (d) $r_{11}=0.1$, $r_{12} = 0.4$, and $r_{22}=0.1$.}
\label{fig:modifier_equilibrium_frequencies}
\end{figure}

Although the equilibrium frequency of the modifier alleles cannot generally be expressed in closed form, the ordering of recombination parameter magnitudes produces predictable trends for modifier evolution, as for stability in Figure~\ref{fig:E1_boundary_comparison}. The following trends are observed in Figure~\ref{fig:modifier_equilibrium_frequencies} for $r=0$ and $s=0.1$. When $r_{12}$ is either the largest or smallest recombination rate (Fig.~\ref{fig:modifier_equilibrium_frequencies}a, b), the direction of modifier evolution depends on the initial modifier frequency. In these cases, $M_1$ will increase when its initial frequency is greater than $0.5$ when $r_{11}=r_{22}$, and decrease if $M_2$ is initially the most frequent modifier allele. When the $M_1M_2$ recombination rate $r_{12}$ is intermediate between $r_{11}$ and $r_{22}$ (Fig.~\ref{fig:modifier_equilibrium_frequencies}c, d), the modifier changes monotonically in frequency: $M_1$ increases when $r_{22} > r_{12} > r_{11}$ and decreases when $r_{11} > r_{12} > r_{22}$. When all three recombination rates are equal, the frequency of the modifier does not change. All trends are qualitatively similar when $r$ is small and non-zero. Modifier evolution depends not only on the relative magnitudes of $r_{11}$, $r_{12}$, and $r_{22}$, but also on the initial genetic composition of the population.

\subsection{Relation to the Fixed-Recombination Model)}
When the modifier is fixed, all individuals experience the same recombination rate and the model reduces to the corresponding two-locus system, where the low fitness equilibrium \(A_1B_1\) is locally stable when

\[
r_{11}>\frac{s}{1+s},
\]

and unstable otherwise. Stability is therefore determined by a single threshold condition: recombination must be sufficiently strong relative to selection to prevent invasion by the fittest haplotype \(A_2B_2\).

With a segregating modifier locus, this condition is no longer sufficient. Rare \(A_2B_2\) haplotypes can arise on different modifier backgrounds and therefore experience different recombination rates. Consequently, stability is determined by the dominant eigenvalue of a \(2\times2\) invasion matrix rather than by a single recombination parameter. For comparison with the chromosomal fusion model of Wirtz et al. \cite{wirtz_impacts_2026}, we write

\[
r_{11}=c,\qquad
r_{12}=r_h,\qquad
r_{22}=r_f,
\]

where \(c\), \(r_h\), and \(r_f\) denote recombination among unfused, mixed, and fused backgrounds, respectively.

\paragraph{Maximal background coupling (\(r=0\)).} When \(r=0\), rare \(A_2B_2\) haplotypes do not move between modifier genotypes. The invasion matrix becomes diagonal, and local stability requires

\[
\alpha c+(1-\alpha)r_h>\frac{s}{1+s},
\]

and

\[
(1-\alpha)r_f+\alpha r_h>\frac{s}{1+s}.
\]

In this limit, the modifier introduces heterogeneity in recombination rates, but it does not generate any interaction between modifier backgrounds. The stability problem decomposes into two independent components, each analogous to the two-locus model.

\paragraph{No background coupling (\(r=\tfrac12\)).} When \(r=\tfrac12\), modifier backgrounds exchange haplotypes at the highest possible rate. Stability now requires

\[
X(\alpha)>0,
\qquad
Y(\alpha)>0,
\qquad
X(\alpha)Y(\alpha)>Z(\alpha),
\]

where

\[
X(\alpha)
=
\alpha c
+(1-\alpha)\frac{1+r_h}{2}
-\frac{s}{1+s},
\]

\[
Y(\alpha)
=
(1-\alpha)r_f
+\alpha\frac{1+r_h}{2}
-\frac{s}{1+s},
\]

and

\[
Z(\alpha)
=
\frac14\alpha(1-\alpha)(1-r_h)^2.
\]

The first two inequalities are direct extensions of the fixed-recombination stability condition to each modifier background. The third condition is qualitatively different. It arises from recombination-mediated exchange of rare \(A_2B_2\) haplotypes between modifier alleles and has no analogue in the fixed-recombination model.

As a result, stability is no longer determined solely by the recombination rates experienced within each background. Even when both backgrounds individually satisfy the stability condition, exchange between backgrounds may destabilize the low-fitness equilibrium.

The comparison highlights the main limitation of a fixed-recombination framework. A single recombination parameter captures only the average rate at which favorable haplotypes are disrupted. It cannot distinguish populations that share the same average recombination rate but differ in how recombination is distributed across genetic backgrounds. Nor can it account for the movement of haplotypes between those backgrounds. Once recombination becomes genetically variable, stability depends not only on the magnitude of recombination, but also on its distribution across modifier alleles and on the rate at which those alleles exchange haplotypes. 

\pagebreak 
\section{Discussion}
The present model extends the framework of Wirtz et al. (2026) by allowing recombination to vary genetically rather than treating it as a fixed parameter. The two models share the same selected-locus structure of \cite{feldman1971equilibrium}. Both consider evolution on a reciprocal-sign-epistasis landscape in which the low-fitness haplotype \(A_1B_1\) and the high-fitness haplotype \(A_2B_2\) are separated by the deleterious intermediates \(A_1B_2\) and \(A_2B_1\). In both models, selection favors \(A_2B_2\) once it is present, whereas recombination can break apart the multilocus association required to maintain it. Thus the core mechanism is the same: evolution depends on the balance between selection for \(A_2B_2\) and recombination-mediated disruption of favorable haplotypes.

At the selected loci, the model of Wirtz et al. is a special case of the present framework. If the population is fixed for one modifier allele, recombination between the selected loci is fixed. The three-locus model then reduces to the corresponding two-locus selection--recombination system. For example, when the population is fixed for modifier allele \(M_1\), the selected-locus recombination rate is \(r_{11}\). Identifying \(r_{11}=c\) recovers the selected-locus dynamics of Wirtz et al., provided that \(c\) denotes the same recombination rate between the selected loci.

The main difference is how recombination is represented. In Wirtz et al., the recombination rate is specified in advance and does not evolve. Their analysis therefore asks how different fixed recombination regimes affect adaptation. In the present model, a modifier locus determines the recombination rates between selected haplotypes. The state space expands from four selected haplotypes to eight haplotypes, and recombination becomes part of the evolving genetic state. This extension makes it possible to study not only the effect of recombination on adaptation, but also how genetic variation in recombination changes evolutionary trajectories and stability of fixations.

This distinction has important consequences. In the two-locus model, the selected-locus state and a single recombination parameter determine the local dynamics. In the modifier model, the same selected-locus state can have different dynamics depending on how selected haplotypes are distributed across modifier alleles. For example, the stability of the \(A_1B_1\) fixation set \(\mathcal E_1(\alpha)\) depends on the modifier composition \(\alpha\), the recombination rates \(r_{11}, r_{12}, r_{22}\), and recombination between the modifier and selected loci. Thus selected-haplotype frequencies alone are generally not sufficient to determine stability.

Although both our model and that of Wirtz et al. involve variable recombination rates, the underlying mechanism is different in each. In Wirtz et al., heterogeneous recombination arises from alternative chromosomal architectures, namely genetic fusions. In the modifier model, heterogeneous recombination arises from segregating modifier alleles; conditions for the stability of $A_1B_1$ fixation can be defined exactly in terms of all four relevant recombination rates. Although these mechanisms differ biologically, they are mathematically similar in that both generate genetic backgrounds with different capacities to preserve or disrupt associations between loci under selection.

Our results also provide a more direct interpretation of how recombination influences the eco-evolutionary outcomes described by Wirtz et al. In that model, persistence near the low-fitness haplotype ($A_1B_1$) is associated with prolonged periods of reduced population growth and an increased risk of population decline. In the present model, the local stability of the corresponding fixation set $\mathcal E_1$ depends on the modifier composition $\alpha$ and on the recombination parameters $r_{11}$, $r_{12}$, $r_{22}$, and $r$. Changes in these recombination fractions can therefore alter whether $\mathcal E_1$ is locally unstable or bistable with the high-fitness fixation set $\mathcal E_4$.

In particular, the relative magnitudes of $r_{11}$, $r_{12}$, and $r_{22}$ determine how modifier composition affects the invasion condition $(1+s)\rho\left(\mathbf A_1(\alpha)\right)<1$. For some parameter combinations, increasing the frequency of a modifier allele stabilizes ($\mathcal E_1$), enlarging the range of states from which the population can remain trapped near the low-fitness haplotype ($A_1B_1$). For other combinations, the same change destabilizes ($\mathcal E_1$), facilitating escape toward fixation of the fitter haplotype ($A_2B_2$). Thus recombination modifiers influence evolution not only by changing the rate at which favorable multilocus combinations are assembled, but also by changing the stability properties of the low-fitness state itself.

 The parameters $r_{11}$, $r_{12}$, and $r_{22}$ in our model correspond to the parameters $c$, $r_h$, and $r_f$ in Wirtz et al., respectively, if $M_1$ and $M_2$ are interpreted as unfused and fused chromosomal arrangements. The parameter $r$, which controls recombination between the modifier locus $M$ and selected locus $B$, has no direct analog in Wirtz et al., where chromosomes are either fused or unfused; thus the correspondence is obtained by setting $r=0$. Under this mapping, our results imply that stability of the low-fitness haplotype $A_1B_1$ requires $\alpha c + (1-\alpha) r_h > s/(1+s)$ and $(1-\alpha)r_f + \alpha r_h > s/(1+s)$, where $\alpha$ is the equilibrium frequency of the unfused karyotype. These conditions provide analytical conditions for which recombination heterogeneity either allows persistence in low-fitness states or facilitates escape toward higher-fitness genotypes. Thus, our results complement the population recovery and valley-crossing dynamics described by Wirtz et al. by identifying explicit stability boundaries for the low-fitness state.

The two models therefore lead to a common biological conclusion. The existent of fitness valley depends on the balance between selection and recombination, and the possibility of demographic recovery is closely tied to whether populations can escape from low-fitness genetic states. The modifier model extends this conclusion by showing that the stability of those states is itself an evolving property of the genetic system. Genetic variation affecting recombination can either reinforce persistence near low-fitness haplotypes, potentially prolonging demographic decline, or destabilize those states and promote movement toward the high-fitness genotype associated with population recovery.

\bibliography{references}

\end{document}